\documentclass[aps,prb,showpacs,twocolumn,citeautoscript,superscriptaddress]{revtex4}
\usepackage{amsmath}
\usepackage{graphicx}
\usepackage{dcolumn}
\usepackage{float}

\begin{document}
\title{Valence fluctuation and magnetic ordering in EuNi$_2$(P$_{1-x}$Ge$_x$)$_2$ single crystals}
\author{U. B. Paramanik}
\affiliation{Department of Physics, Indian Institute of Technology, Kanpur 208016, India}
\author{A. Bar}
\affiliation{Department of Physics, Indian Institute of Technology, Kanpur 208016, India}
\author{Debarchan Das}
\affiliation{Department of Physics, Indian Institute of Technology, Kanpur 208016, India}
\author{N. Caroca-Canales}
\affiliation{Max-Planck Institute for Chemical Physics of Solids, 01187 Dresden, Germany}
\author{R. Prasad}
\affiliation{Department of Physics, Indian Institute of Technology, Kanpur 208016, India}
\author{C. Geibel}
\affiliation{Max-Planck Institute for Chemical Physics of Solids, 01187 Dresden, Germany}
\author{Z. Hossain}
\email{zakir@iitk.ac.in}
\affiliation{Department of Physics, Indian Institute of Technology, Kanpur 208016, India}
\affiliation{Max-Planck Institute for Chemical Physics of Solids, 01187 Dresden, Germany}

\date{\today}

\begin{abstract}

Unusual phases and phase transitions are seen at the magnetic-nonmagnetic boundary in Ce, Eu and Yb-based compounds. EuNi$_2$P$_{2}$ is a very unusual valence fluctuating Eu system, because at low temperatures the Eu valence stays close to 2.5 instead of approaching an integer value. Eu valence and thus the magnetic property in this system can be tuned by Ge substitution in P site as EuNi$_2$Ge$_{2}$ is known to exhibit antiferromagnetc (AFM) ordering of divalent Eu moments with $T_N$ = 30 K. We have grown EuNi$_2$(P$_{1-x}$Ge$_x$)$_2$ (0.0 $\leq$ $x$ $\leq$ 0.5) single crystals and studied their magnetic, thermodynamic and transport properties. Increasing Ge doping to $x >$ 0.4 results in a well-defined AFM ordered state with $T_N$ = 12 K for $x$ = 0.5. Moreover, the reduced value of magnetic entropy for $x$ = 0.5 at $T_N$ suggests the presence of valance fluctuation/ Kondo effect in this compound. Interestingly, the specific heat exhibits an enhanced Sommerfeld coefficient upon Ge doping. Subsequently, electronic structure calculations lead to a non-integral valence in EuNi$_2$P$_{2}$ but a stable divalent Eu state in EuNi$_2$Ge$_{2}$ which is in good agreement with experimental results.

\end{abstract}


\maketitle

\section{INTRODUCTION}

Recent decades abound with research on the strongly correlated systems having instabilities of inner $f$-electron shells, mainly due to a wide range of fascinating phenomena discovered in those systems, such as ``heavy-fermion (HF) superconductivity", Kondo insulators, valence fluctuation (VF), non-Fermi liquid (NFL) behavior, quantum criticality etc.\cite{Brandtc5} Moreover, long-range magnetic ordering in the system can be destabilized by tuning external or internal (alloying) pressure or by application of external magnetic field leading the system to the borderline between magnetically ordered and nonmagnetic ground states. The precarious point of instability between two stable phases of matter is called the `quantum critical point' (QCP) which is the playground of new interesting physics. In addition to that, valence instability which is  prone to occur near the middle of the lanthanide series, is a key factor which governs the ground state and can also lead the system to QCP. While most experimental and theoretical efforts have been focused on the Ce and Yb elements,\cite{Monthouxc5} the Eu compounds are relatively unexplored.

Similar to those of Ce and Yb based systems, one can tune the electronic ground state of Eu compounds. However, the evolution of the valence $v$ of Eu-based systems as a function of temperature $T$ and composition $x$ or pressure $p$ differs strongly from the phase diagrams for Ce or Yb systems. In Ce or Yb systems, increasing the hybridization results in a smooth and continuous evolution of $v$ as a function of $p$ or $x$ and the magnetic ordering temperature $T_N$ decreases continuously to zero value at a quantum critical point.\cite{Doniachc5, Lacroixc5} On the other hand, for Eu-based systems, increasing hybridization leads to a slight increase in $T_N$ rather than decreasing. At a critical value an abrupt transition to an almost trivalent Eu ground state occurs.

The characteristic phase diagram for Eu systems was first established in the system Eu(Pd$_{1-x}$Au$_x$)$_2$Si$_2$ \cite{Segrec5}, and later was confirmed in Eu(Pt$_{1-x}$Ni$_x$)$_2$Si$_2$ \cite{Mitsudac5}. Ni substitution for Pt in EuPt$_2$Si$_2$ acts as a chemical pressure in the system and suppresses the antiferromagnetism (AFM) completely. More recently, the substitution of Si in EuCu$_2$Ge$_2$ was shown to be of particular interest because the rare-earth ion configuration changes from nearly integral valence Eu$^{2+}(4f^7)$ in antiferromagnetic EuCu$_2$Ge$_2$ to a valence fluctuating state in EuCu$_2$Si$_2$. A strong electron mass enhancements was reported in the compounds near the crossover from magnetic to non-magnetic state \cite{Hossainc5}. One striking feature of this system is the coexistence of the magnetically ordered phase with valence fluctuations.\cite{Hossainc5, Alekseev} We present the the crystal growth of EuNi$_2$(P$_{1-x}$Ge$_x$)$_2$ and study the evolution from valence-fluctuating EuNi$_2$P$_2$ to divalent Eu upon increasing Ge doping. One advantage of substituting Ge for P instead of e.g. Fe for Ni is that both EuNi$_2$P$_2$ and EuNi$_2${Ge}$_2$ crystallize in the so called ``collapsed structure", while substituting Fe for Ni results in a change from collapsed to non-collapsed structure, for which one can expect strong disorder effect.

The compound EuNi$_2$P$_2$ exhibits valence fluctuating behavior and possess a gradual change of the Eu valence with decreasing temperature.\cite{Hiranakac5, Guritanu} The temperature dependence of Eu valence was also probed by Mo\"{o}ssbauer spectroscopy where isomer shift gradually moves with decreasing temperature from -6.4~mm/s at room temperature to -4.4mm/s at 4.2~K. \cite{Nagarajanc5} Photoemission spectroscopy study on high quality single crystals of EuNi$_2$P$_2$ revealed the presence of heavy bands in the compound \cite{Geibelc5}. On the other hand, isostructural EuNi$_2$Ge$_2$ \cite{Zahirulc5} exhibits a stable divalent Eu state which orders antiferromagnetically at 30.8~K. Anticipating the fact that Ge substitution for P in EuNi$_2$P$_2$ may change the magnetic ground state of Eu and the system may exhibit quantum critical behavior at the magnetic non-magnetic borderline, we have carried out a systematic study of EuNi$_2$(P$_{1-x}$Ge$_x$)$_2$ system.

\section{METHODS}

We have grown the single crystals of EuNi$_2$(P$_{1-x}$Ge$_x$)$_2$ using Sn-flux. The starting elements Eu, Ni, (P$_{1-x}$Ge$_x$) and Sn were taken in the ratio of 2:1:1.2:5 and the mixture was placed in an alumina crucible and subsequently sealed in an evacuated quartz ampoule. The sealed quartz ampoule was slowly heated to 1050$^\circ$C and held at that temperature for 20 hours for proper mixing. It was then cooled down at a rate of 2$^\circ$C/h down to 500$^\circ$C and then rapidly brought down to room temperature in two hours. Single crystals were removed by etching the Sn flux using dilute HCl. We obtained plate-like single crystals with different sizes depending on the Ge concentration varied from 4$\times$4$\times$0.2~mm$^3$ to 2$\times$2$\times$0.1~mm$^3$. The samples were characterized by powder x-ray diffraction with Cu-$K_\alpha$ radiation to determine the phase purity and crystal structure. We studied the compositional homogeneity for EuNi$_2$(P$_{1-x}$Ge$_x$)$_2$  samples using Scanning electron microscope (SEM) equipped with energy dispersive x-ray (EDX) analysis.

\section{RESULTS AND DISCUSSION}

\subsection*{\label{ExpDetails} A. Crystal Structure}

\begin{figure}
\includegraphics[width=8.7cm, keepaspectratio]{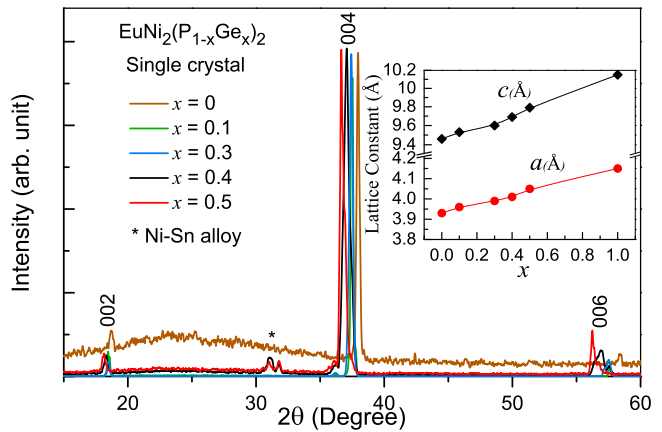}
\caption{\label{fig:EuNi2GeP-xrd} (Color online) x-ray diffraction patterns for EuNi$_2$(P$_{1-x}$Ge$_x$)$_2$ ($x$ = 0, 0.1, 0.3, 0.4 and 0.5) plate-like single crystals. Inset shows the calculated lattice constants.}
\end{figure}

Figure~\ref{fig:EuNi2GeP-xrd} shows the x-ray diffraction pattern of EuNi$_2$(P$_{1-x}$Ge$_x$)$_2$ ($x$ = 0, 0.1, 0.3, 0.4 and 0.5) plate like single crystals. We observe only the $(00l)$ reflections which indicate that the $c$-axis is perpendicular to the crystal plane. For the crystals of above 30\% Ge concentration, two small extra x-ray peaks appear at around 31$^\circ$ which are coming from tiny unreacted Ni-Sn phase attached to the surface of the crystals. SEM images for the 40\% and 50\% Ge doped crystals also reveal a few scattered tiny unreacted Ni-Sn alloy flakes attached to the surface of the crystals. Upon increasing the Ge-concentration, the $(00l)$ peaks monotonically shifts toward a lower 2$\theta$ angle, indicating a continuous increase of the $c$ axis lattice parameter with increasing $x$, suggesting less hybridization between the Eu-layer and the Ni-P,Ge layer. The other lattice parameter $a$ = $b$ was calculated from the powdered x-ray diffraction on the crushed single crystals. The analysis of powder x-ray diffraction data revealed that all the compounds EuNi$_2$(P$_{1-x}$Ge$_x$)$_2$ crystallize in ThCr$_2$Si$_2$-type tetragonal crystal structure (space group \textit{I4/mmm}). The obtained lattice parameters for EuNi$_2$P$_2$ are in close agreement with the values as reported in literature \cite{Hiranakac5}. The $c$-axis lattice parameter increases anisotropically with increasing Ge doping concentration.

%Table II
\begin{table}
\centering
\caption [Lattice parameters $a${\AA}, $c${\AA}, $c/a$ ratio and unit-cell volume $V${\AA$^3$} of ThCr$_{2}$Si$_{2}$-type tetragonal system Eu(Fe$_{1-x}$Ir$_{x}$)$_{2}$As$_{2}$ ($x$ = 0, 0.05, 0.11 and 0.14)]{Lattice parameters $a$({\AA}), $c$({\AA}), $c/a$ ratio and unit-cell volume $V$({\AA$^3$}) of ThCr$_{2}$Si$_{2}$-type tetragonal system EuNi$_2$(P$_{1-x}$Ge$_x$)$_2$ ($x$ = 0, 0.1, 0.3, 0.4, 0.5 and 1.0).}
\label{table:EuNiGeP-table1}
\vskip .5cm
\addtolength{\tabcolsep}{+5pt}
\begin{tabular}{c c c  c c}
\hline
\hline
$x$ & $a$ ({\AA})  &$c$ ({\AA})	& $c/a$			& $V$ ({\AA$^3$})\\[0.5ex]
\hline
0 	 & 3.93(8)	&9.46(5)	& 2.40			& 146.84(4)\\[1ex]

0.1 & 3.96(9)	&9.54(3)		& 2.41		& 149.6(5)\\[1ex]

0.3 & 3.99(19)	&9.60(5)	&  2.41			& 152.9(11)\\[1ex]

0.4 & 4.01(14)	&9.69(6)	&  2.41			& 155.8(9) \\[1ex]

0.5 & 4.05(4)	&9.82(8)	& 2.42			& 161.1(3)\\[1ex]

1.0 & 4.14(9)	&10.15(6)	& 2.45			& 172.2(4)\\
\hline

\end{tabular}
\end{table}

\subsection*{\label{ExpDetails} B. Magnetization}

\begin{figure}
\includegraphics[width=8.5cm, keepaspectratio]{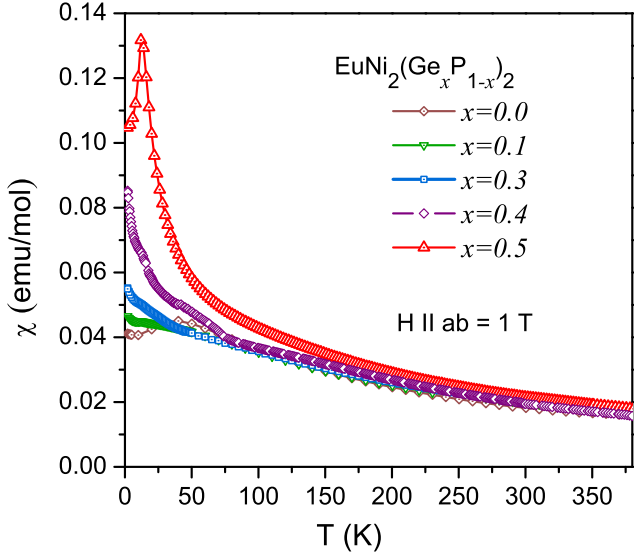}
\caption{\label{fig:EuNi2GeP-MT} (Color online) Temperature dependence of the dc magnetic susceptibility $\chi(T)$ for EuNi$_2$(P$_{1-x}$Ge$_x$)$_2$: 0 $\leq$ $x$ $\leq$ 0.5.}
\end{figure}

\begin{figure}
\includegraphics[width=8.5cm, keepaspectratio]{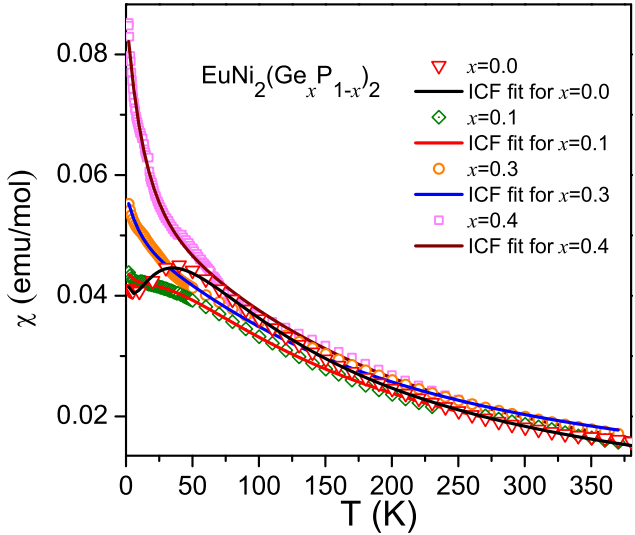}
\caption{\label{fig:EuNiGeP-ICF} (Color online) Temperature dependence of the magnetic susceptibility of EuNi$_2$(P$_{1-x}$Ge$_x$)$_2$ ($x$ = 0, 0.1, 0.3 and 0.4). The solid lines through the magnetic susceptibility data show the fit to the two-level ionic interconfiguration fluctuations (ICF) model.}
\end{figure}

The temperature dependent magnetic susceptibilities for EuNi$_2$(P$_{1-x}$Ge$_x$)$_2$ (0 $\leq$ $x$ $\leq$ 0.5) are shown in Figure~\ref{fig:EuNi2GeP-MT}. At higher temperatures the susceptibility follows modified Curie-Weiss (CW) law: $\chi(T)= \chi_0 + C/(T - \theta)$, where $C$ is the Curie constant and $\theta$ is the Curie-Weiss temperature. The calculated effective magnetic moment $\mu_{eff}$ = 6.9 $\mu_{B}$/Eu for EuNi$_{2}$P$_{2}$ lies in between the values for Eu$^{2+}$ and Eu$^{3+}$. The effective moments are slightly increased with increasing Ge content up to $x$ = 0.4. However, the values are still smaller than $\mu_{eff}$ = 7.94 $\mu_{B}$/Eu for a stable divalent Eu$^{2+}$ ion. The shape of the magnetic susceptibilities of EuNi$_2$(P$_{1-x}$Ge$_x$)$_2$ with 0 $\leq x \leq$ 0.4 (Figure~\ref{fig:EuNiGeP-ICF}) are typical of those for valence fluctuating Ce compounds \cite{Layekc5, Rojasc5, Kowalczykc5, Mazumdarc5} where sometimes the magnetic susceptibility shows a curie tail at lower temperatures \cite{Mazumdarc5, Allenoc5}, which arises due to presence of magnetic impurity.

Sales and Wohlleben developed a model (interconfiguration fluctuation model, ICF) addressing this behavior. \cite{Salesc5}  The model says that the overall magnetic susceptibility can be described by the sum of three different parts, a valence fluctuation part, a part for stable Eu$^{2+}$ (including the trace amounts of possible impurities) and a temperature independent part.  Thus, the temperature dependence of the susceptibility is given by

\begin{equation}\begin{split}
\chi(T) & = \left(\frac{N}{3k_{B}}\right) \left[\frac{\mu_{n}^2 \nu(T)+ \mu_{n-1}^2\{1-\nu(T)\}}{T^*}\right]\\
\quad & + f\frac{C}{T-\theta}+\chi_{0}
\label{eq:chi5}
\end{split}\end{equation}

with

\begin{equation}
\nu(T) = \frac{2J_{n}+1}{(2J_{n}+1) + (2J_{n-1}+1)exp(-E_{ex}/K_{B}T^*)}
\label{eq:chi2}
\end{equation}

and
\begin{equation}
T^* = \left[T+T_{sf}\right]
\label{eq:chi3}
\end{equation}

\noindent where $\mu_{n}$ and $\mu_{n-1}$ are the effective moments in $4f^n$ and $4f^{(n-1)}$  states, $(2J_{n}+1)$ and $(2J_{n-1}+1)$ are the degeneracies of the corresponding energy states E$_n$ and E$_{n-1}$. In these expressions, E$_{ex}$ is interconfigurational excitation energy which is equal to (E$_{n}$-E$_{n-1})$, where Eu$^{3+}$ ($E_n$, $J_n$ = 0 and $\mu$ = 0~$\mu_{B}$) state is the ground state and Eu$^{2+}$ ($E_{n-1}$, $J_{n-1}$ = $\frac{7}{2}$ and $\mu$ = 7.94~$\mu_{B}$) is the excited state.  $T_{sf}$  is the spin fluctuation temperature associated with the valence fluctuation, $f$ is fraction of stable Eu$^{2+}$ ions responsible for the CW behavior. Thus the final equation is

%Table II
\begin{table}
\centering
\caption [Crystallographic and Magnetic parameters of doped EuNi$_2$P$_2$ compounds]{Effective paramagnetic moments $\mu_{eff}$($\mu_{B}$), interconfigurational excitation energy E$_{ex}$(K), spin fluctuation temperature $T_{sf}$  and antiferromagnetic ordering temperature $T_N$(K) of EuNi$_2$(P$_{1-x}$Ge$_x$)$_2$.}
\label{table:EuNiGeP-table1}
\vskip .5cm
\addtolength{\tabcolsep}{+5pt}
\begin{tabular}{c c c c c}
\hline
\hline
$x$ & $\mu_{eff}$($\mu_{B}$) & $E_{ex}/K_B(K)$  & $T_{sf}(K)$ & $T_{N}$(K) \\[0.5ex]
\hline
0 	&  6.9	&192 & 53 &   - \\[1ex]

0.1 &  7.2	& 227 & 77 &  -\\[1ex]

0.3 &  7.7	&274 & 91 &  - \\[1ex]

0.4 &  7.8	&333 & 81 &  - \\[1ex]

0.5 &  8.0	&- & - &  12\\[1ex]

1.0 & 8.0 & - & - &  30.8 \\
\hline
\end{tabular}
\end{table}

\begin{equation}\begin{split}
\chi(T) &= \left(\frac{N}{3k_{B}}\right)\left[\frac{(7.94\mu_{B})^2\{1-\nu(T)\}}{T^*}\right]\\
 \quad & + f\frac{C}{T-\theta}+\chi_{0}
\label{eq:chi4}
\end{split}\end{equation}

with

\begin{equation}
\nu(T) = \frac{ 1}{1 + 8exp(-E_{ex}/K_{B}T^*)}
\label{eq:chi5}
\end{equation}

\noindent The solid line through the magnetic susceptibility data in Fig.~\ref{fig:EuNiGeP-ICF} is the fit to the equation 4. The values of parameters obtained from the least square fits are listed in Table~\ref{table:EuNiGeP-table1}. The values of $f$ (0.02 to 0.2), E$_{ex}$ and $T_{sf}$ are in the range of typical values found for other valence-fluctuating system Eu$_{0.4}$La$_{0.6}$Pd$_3$ (E$_{ex}$ = 418 K and $T_{sf}$ = 164 K).\cite{Pandeyc5} However, the values of excitation energies and spin fluctuation temperatures are found to increase with increasing $x$ which is inconsistent with our finding that the system attains a stable magnetically ordered state with increasing Ge-doping. This inconsistency might arise due to the fact that for EuNi$_2$P$_2$ one needs to include on-site Coulomb repulsion ($U_{fd}$) between $4f$ and $5d$ electron, i.e the Falikov Kimball term along with hybridization ($V_f$) between $4f$ and valence electrons.  However, such analysis is not possible at this juncture as evolution of $U_{fd}$ with $x$ is not known.

\begin{figure}
\includegraphics[width=8.8cm, keepaspectratio]{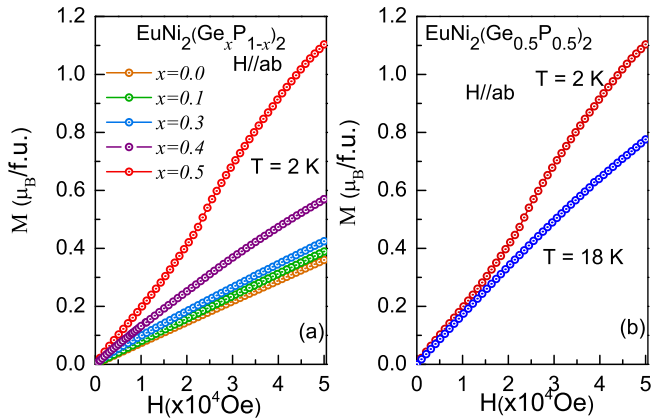}
\caption{\label{fig:EuNi2GeP-MH} (Color online) (a) Isothermal magnetization M(H) of EuNi$_2$(P$_{1-x}$Ge$_x$)$_2$ at 2~K for H$\parallel$$ab$. (b) Isothermal magnetization $M(H)$ of EuNi$_2$(P$_{0.5}$Ge$_{0.5}$)$_2$ at 2~K and 18~K.}
\end{figure}

For  50\% Ge doped sample the effective paramagnetic moment is very close to that of a free Eu$^{2+}$ ion (4$f^7$: $S$ = 7/2, $L$ = 0, and $J$ = 7/2). Also, a susceptibility peak corresponding to the transition to an antiferromagnetic state occurs at $T_N$  = 12~K for x = 0.5, confirming that the system evolves from a valence fluctuating state to a divalent state of Eu upon increasing Ge concentration. To further elucidate the magnetic properties, we carried out the high-field isothermal magnetization measurements at temperatures 2~K and 18~K with applied field $H$ parallel to the $ab$ plane of the crystals. As shown in Figure~\ref{fig:EuNi2GeP-MH}(a), for lower doping concentration of Ge up to 40\%, the magnetization increases linearly with the field. Also, the slope of the lines becomes steeper as we increase the Ge doping concentration, indicating the increase in paramagnetic moment. For EuNi$_2$(P$_{0.5}$Ge$_{0.5}$)$_2$, the magnetization varies almost linearly with the field in the range 0 Oe $\leq$ H $\leq$ 20~kOe, accompanied by a change in slope at $H_c \approx$ 20~kOe, indicating a spin flop-like transition which is usual for antiferromagnets. As shown in Figure~\ref{fig:EuNi2GeP-MH}(b), the $M(H)$ isotherm for EuNi$_2$(P$_{0.5}$Ge$_{0.5}$)$_2$ exhibits spin flop transition for $T$$\leq$$T_N$ while $M$ varies linearly with $H$ above the magnetic ordering temperature.

\subsection*{\label{ExpDetails} C. Specific heat}

\begin{figure}
\includegraphics[width=8.0cm, keepaspectratio]{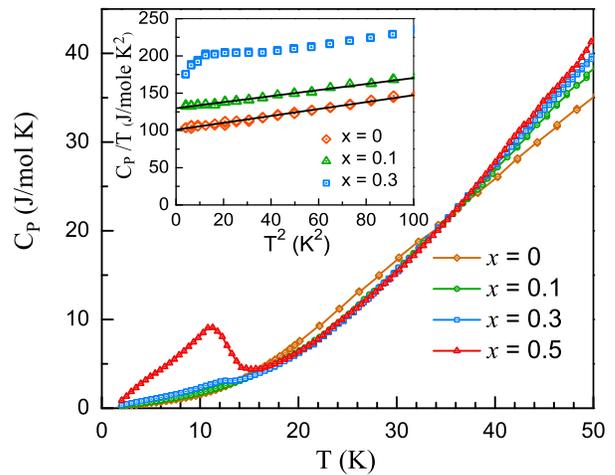}
\caption{\label{fig:EuNi2GeP-CP} (Color online) Specific-heat C$_P$($T)$ as a function of temperature $T$ for EuNi$_2$(P$_{1-x}$Ge$_x$)$_2$: 0 $\leq$ $x$ $\leq$ 0.5. Inset: Low-temperature specific heat data divided by temperature $C_P$/$T$ versus squared temperature. The solid lines represent linear fits to the data in the range 1 $\leq T^2 \leq$ 100 with C$_P$/$T$ = $\gamma$ + $\beta T^2$.}
\end{figure}

The temperature dependencies of the specific heat $C_P(T)$ of the series are displayed in Fig.~\ref{fig:EuNi2GeP-CP}. For $x$ $\leq$ 0.3, the specific heat exhibits no anomaly down to 2~K and hence confirms the absence of any magnetic ordering. When we plot the $C_P(T)$/$T$ versus $T^2$ data (inset of Fig.~\ref{fig:EuNi2GeP-CP}), a linear region is observed below 10~K for $x$ = 0 and 0.1. The low temperature data of $C_P(T)$/$T$ vs $T^2$ below 10~K are fitted to the equation
\begin{equation}
C_P(T)/T = \gamma  + \beta T^{2}
\label{eq:C}
\end{equation}
where $\gamma T$ term is the contribution of the conduction electrons and the $\beta$ term reflects the phonon contribution to the specific heat. The fit to the data yields the value of $\gamma = 98$~mJ/mol\,K$^{2}$ and $\beta =4.6\times10^{-4}$~J/mol\,K$^{4}$\@ for EuNi$_2$P$_2$. The obtained value of Sommerfeld coefficient for EuNi$_2$P$_2$ is in good agreement with the literature \cite{Hiranakac5}. Further, we estimate the Debye temperature $\Theta_{\rm D}$ for EuNi$_2$P$_2$ from $\beta$ using the relation \cite{Kittelc5}
\begin{equation}
\Theta_{\rm D} = \left( \frac{12 \pi^{4} n R}{5 \beta} \right)^{1/3}
 \label{eq:Debye-Temp}
\end{equation}
\noindent where $R$ is molar gas constant and $n = 5$ is the number of atoms per formula unit (f.u.)\@. We obtain Debye temperature $\Theta_{\rm D}$ = 276~K\@ for EuNi$_2$P$_2$.
For a 10\% Ge substituted sample EuNi$_2$(P$_{0.9}$Ge$_{0.1}$)$_2$, we observe an enhanced Sommerfeld coefficient as $\gamma = 130$~mJ/mol\,K$^{2}$, indicating an effective mass enhancement for the conduction electrons in the valence-fluctuating regime. For $x = 0.3$ the $\beta$ coefficient remains same ($4.6\times10^{-4}$~J/mol\,K$^{4}$) but the Sommerfeld coefficient increases significantly ($\gamma = 185$~mJ/mol\,K$^{2}$). However, the estimation of the Sommerfeld coefficient ($\gamma$) for $x = 0.3$ may not be reliable as a slight deviation is seen in the $C_P(T)$/$T$ vs $T^2$ curve below 2.25~K, the origin of which is not yet understood. Although the heavy fermion behavior has been observed in many Ce-based compounds, there are very few examples of Eu-based heavy fermion compounds.\cite{Hossainc5} For $x = 0.5$ a broadened mean field-type transition is observed at $T_N$ = 12~K corresponding to the antiferromagnetic transition and hence confirms the intrinsic and bulk nature of the magnetic ordering.

\begin{figure}
\includegraphics[width=8.7cm, keepaspectratio]{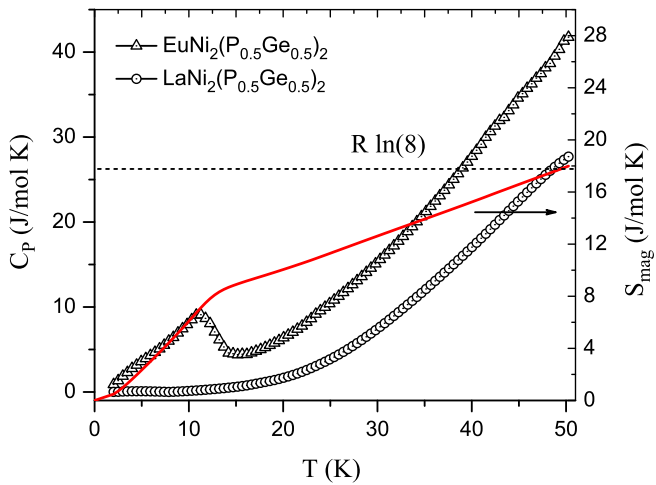}
\caption{\label{fig:EuNi2GeP-Sm} (Color online) Specific-heat $C_P$($T)$ of EuNi$_2$(P$_{0.5}$Ge$_{0.5}$)$_2$ and LaNi$_2$(P$_{0.5}$Ge$_{0.5}$)$_2$, magnetic part of the entropy S$_{mag}$($T)$.}
\end{figure}

In order to isolate the magnetic contribution to the specific-heat $C_{mag}(T)$, $C_{P}(T)$ data for the isostructural nonmagnetic reference compound LaNi$_2$(P$_{0.5}$Ge$_{0.5}$)$_2$ was used from Reference[28]. Figure~\ref{fig:EuNi2GeP-Sm} shows the temperature dependence of heat capacity of EuNi$_2$(P$_{0.5}$Ge$_{0.5}$)$_2$, together with the heat capacity of nonmagnetic reference polycrystalline LaNi$_2$(P$_{0.5}$Ge$_{0.5}$)$_2$. The magnetic part of the heat capacity $C_{mag} (T)$ was deduced by the usual method of subtracting the heat capacity of LaNi$_2$(P$_{0.5}$Ge$_{0.5}$)$_2$ from that of EuNi$_2$(P$_{0.5}$Ge$_{0.5}$)$_2$ after adjusting the renormalization due to different atomic masses of La and Eu. The magnetic contribution to the entropy $S_{mag}$ was calculated by integrating the $C_{mag}/T$ versus $T$. The magnetic entropy $S_{mag}$ at $T_N$ is coming out to be 8~J/mol\,K which is only 46\% of $Rln8$ = 17.3~J/mol\,K corresponding to the theoretical value for Eu$^{2+}$ moments ($S$ = 7/2). The reduced value of magnetic entropy at $T_N$ suggests the presence of valence fluctuation/ Kondo effect in this compound. The magnetic entropy reaches to the expected theoretical value for divalent Eu moments at around 50~K and it keeps on increasing beyond 50K. This indicates that the lattice contribution to the heat capacity of EuNi$_2$(P$_{0.5}$Ge$_{0.5}$)$_2$ is larger than that of LaNi$_2$(P$_{0.5}$Ge$_{0.5}$)$_2$. This, however, does not affect the conclusion that the magnetic entropy at $T_N$ is much less than expected.

\subsection*{\label{ExpDetails} D. Electrical Resistivity}

\begin{figure}[htb!]
\includegraphics[width=8.7cm, keepaspectratio]{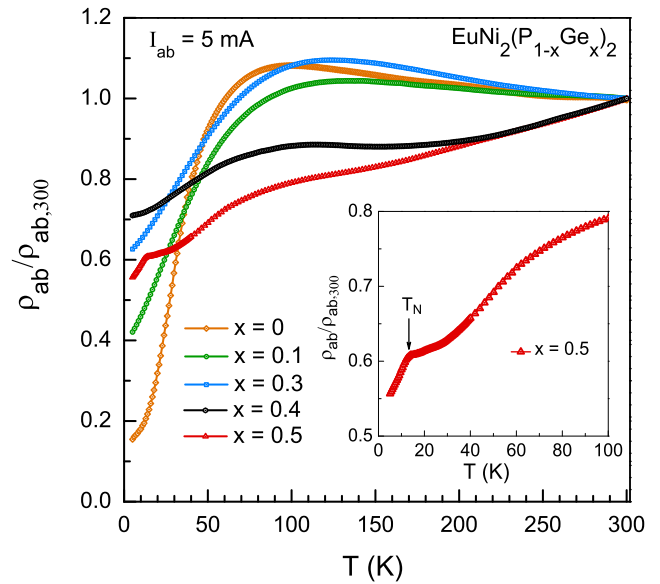}
\caption{\label{fig:EuNi2GeP-RT} (Color online) Temperature dependence of the electrical resistivity for EuNi$_2$(P$_{1-x}$Ge$_x$)$_2$: 0 $\leq$ $x$ $\leq$ 0.5.}
\end{figure}

The temperature dependencies of the electrical resistivity $\rho(T)$ of EuNi$_2$(P$_{1-x}$Ge$_x$)$_2$ with 0 $\leq$ $x$ $\leq$ 0.5 are displayed in Figure~\ref{fig:EuNi2GeP-RT}. For $x$  $\leq$ 0.4, the resistivity curves are reminiscent of that of observed in heavy fermion Kondo lattice systems where $\rho(T)$ increases with decreasing temperature below room temperature, passes through a broad maximum at a temperature related to the Kondo temperature $T_K$ and then decreases at lower temperature due to the onset of coherence. Such resistivity behavior are typical for a heavy fermion Kondo lattice systems \cite{Knoppc5, Schoenesc5} containing Ce, Yb or U but it is not so common in Eu-based compounds. For $x$ = 0.4 the broad maxima around 100~K becomes weaker. For a 50\% Ge doped compound the $\rho(T)$ shows a sudden decrease below $T_N$ = 12~K due to the Eu-moment ordering, accompanied by a broad resistivity hump at higher temperature.The loss of resistance below $T_N$ is due to the loss of magnetic scattering in the AFM ordered state. Similar resistivity behavior is observed in many Ce-based magnetically ordered Kondo lattice system, such as CeNi$_x$Pt$_{1-x}$ \cite{Dasc5} and CePd$_2$(Si$_{1-x}$Ge$_x$)$_2$ \cite{Gignouxc5} where RKKY exchange interactions compete with the Kondo effect.

\subsection*{\label{ExpDetails} E. Electronic Structure calculations}

\begin{figure}[htb!]
\includegraphics[width=8.2cm, keepaspectratio]{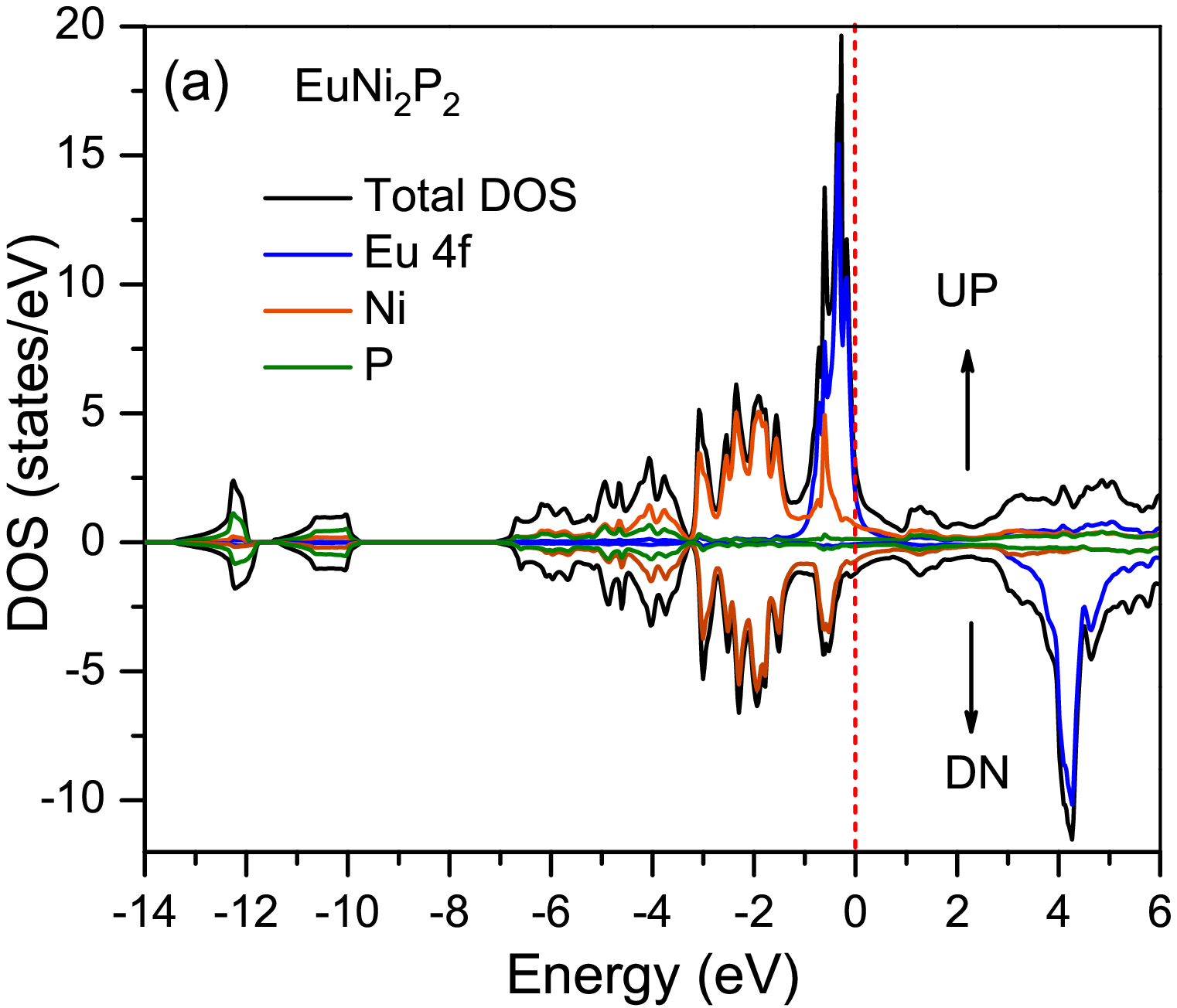}
\includegraphics[width=8.2cm, keepaspectratio]{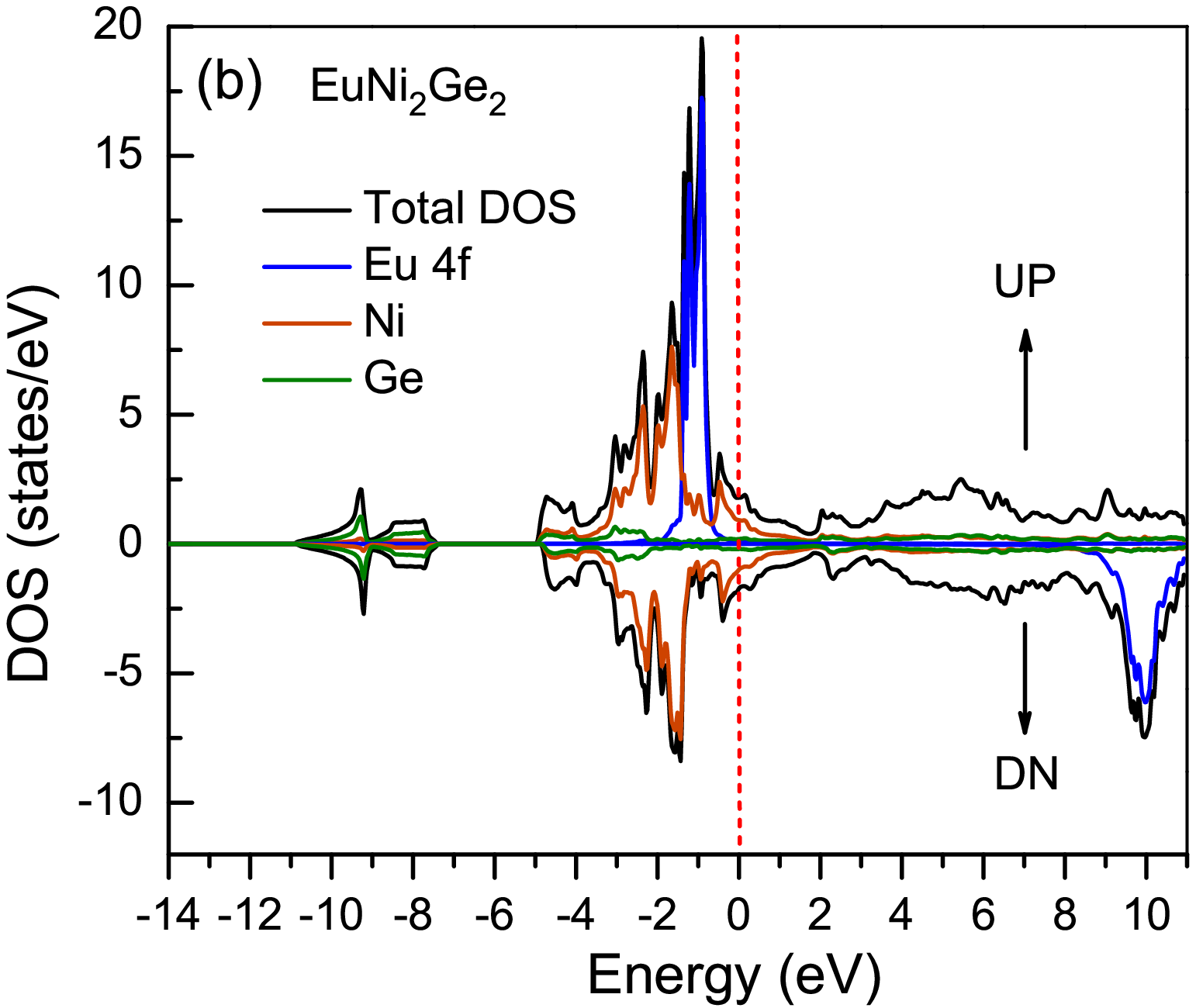}
\caption{\label{fig:EuNi2GeP-DOS} (Color online) The spin-polarized density of states calculated for (a) EuNi$_2$P$_2$ and (b) EuNi$_2$Ge$_2$ with U = 7.4 eV, J = 0.7 eV. The Fermi level (E$_{F}$) corresponds to zero binding energy.}
\end{figure}

In order to gain some information on the electronic state of the compounds EuNi$_{2}$P$_{2}$ and  EuNi$_{2}$Ge$_{2}$ (compounds of two extreme ends in the series EuNi$_2$(P$_{1-x}$Ge$_x$)$_2$) we have carried out the density-functional band structure calculations using the full potential linear augmented plane wave plus local orbitals (FP-LAPW+lo) method as implemented in the WIEN2k code \cite{Blahac5}. The set of plane-wave expansion K$_{MAX}$ was determined as R$_{MT}\times$ K$_{MAX}$ equal to 7 and K mesh used was $10\times10\times10$. The exchange-correlation potential was calculated using the generalized gradient approximation (GGA) as proposed by Pedrew, Burke and Ernzerhof. \cite{Perdewc5}Additionally, to account the on-site Coulomb interaction, we have included a strong Coulomb repulsion in the Eu-$4f$ orbitals on a mean-field level using the GGA+U approximation. For Eu the values of U and $J$ parameters are chosen after Johannes and Pickett \cite{Pickettc5}. In addition, the spin orbit coupling is included with the second variational method in the Eu-$4f$ shell.

The  total and partial DOS  of  EuNi$_{2}$P$_{2}$ and  EuNi$_{2}$Ge$_{2}$  are  shown in  Figure~\ref{fig:EuNi2GeP-DOS}. For our GGA+U calculations, we have chosen a value of U = 7.4 eV for the effective Coulomb repulsion and a value of $J$ = 0.7 eV for the intra-atomic $4f$ exchange interaction on the Eu ion, similar like in EuCo$_{2}$X$_{2}$ (X = Ge, Si) system \cite{Deniszczykc5}. The general shape of the DOSs are very similar to that of EuCo$_{2}$Ge$_{2}$ or EuCo$_{2}$Si$_{2}$ system \cite{Deniszczykc5}. In both the compounds EuNi$_{2}$P$_{2}$ and  EuNi$_{2}$Ge$_{2}$, the deeply lying valence band (below -7.5~eV) is mainly composed of $s$ states of P and Ge atoms and separated by energy gap of 2.7 eV in EuNi$_{2}$P$_{2}$ and of 2.4 eV in EuNi$_{2}$Ge$_{2}$ from the main complex of valence bands close to the fermi level (E$_F$). As can be seen from Fig.~\ref{fig:EuNi2GeP-DOS}, the valence bands below the fermi level are much more extended for EuNi$_{2}$P$_{2}$ (0 to -7 eV) as compared to EuNi$_{2}$Ge$_{2}$ (0 to -5 eV). The extended DOS is associated with stronger $p$-$d$ hybridization in EuNi$_{2}$P$_{2}$. In both the compounds the up and down Eu-$4f$ electrons spins are polarized and well separated. In case of EuNi$_{2}$P$_{2}$, the bottom of the conduction i.e. the Eu-$4f$ majority (up) band lies at the Fermi level. The Eu-$4f$ orbitals hybridized slightly with the Ni-$3d$ orbitals yielding the nonintegral valence Eu$^{2.5+}$ (4$f^{6.5}$) of europium in EuNi$_{2}$P$_{2}$ compound. The calculated DOS is consistent with the photoemission spectroscopy results where the presence of Eu-4$f$ density of states at $E_F$ is significant in EuNi$_{2}$P$_{2}$ \cite{Felserc5}. The magnetic moment comes from the spin polarized Eu-$4f$ states only ($\sim$ 6.5~$\mu_{B}$). In EuNi$_{2}$Ge$_{2}$, the majority spin up channel of Eu-4$f$ states become fully occupied as expected. These strongly localized states are placed at about 1 eV below the fermi level. The $4f$ states of the minority-spin (dn) channel remain unoccupied and are shifted upwards to 10.5~eV. Evaluating the occupation of the $4f$ orbital, EuNi$_{2}$Ge$_{2}$ is found to be in a magnetic $4f^7$ state.

\section{\label{Conclusions} CONCLUSIONS}

We have successfully grown single crystals of EuNi$_2$(P$_{1-x}$Ge$_x$)$_2$ with 0 $\leq$ $x$ $\leq$ 0.5 using Sn-flux. The evolution from valence fluctuating state to magnetically ordered state in EuNi$_2$(P$_{1-x}$Ge$_x$)$_2$ is presented. A comparative study of lattice parameters of EuNi$_2$(P$_{1-x}$Ge$_x$)$_2$ suggests that there is an expansion of unit-cell volume across the series from $x$ = 0 to 1. The expansion of unit-cell volume with increasing $x$ (Ge) causes dehybridization of Eu-$4f$ electron with the conduction electrons which increases the localized character of the $4f$ electron.  The enhancement of localization results in a decrease in Eu valence and becomes divalent in the 50\% Ge doped sample. The weakening of valence fluctuations allows the RKKY exchange to become more effective, leading to a magnetically ordered ground state in EuNi$_2$(P$_{0.5}$Ge$_{0.5}$)$_2$. In order to have a clear picture about the valence state of Eu close to the borderline between magnetic and non-magnetic boundary, photoemission measurements will be worthwhile to study in this system. Furthermore, our results suggest that the antiferromagnetically ordered state of Eu is very stable for $x$ = 1 to $x$ = 0.5 but it vanishes abruptly at $x$ = 0.4 which can be considered as a critical concentration ($x_c$) in EuNi$_2$(P$_{1-x}$Ge$_x$)$_2$ series. In addition, we observe an enhanced Sommerfeld coefficient upon increasing Ge content. Subsequently, we have presented the results of electronic structure calculations for the two systems EuNi$_2$P$_2$ and EuNi$_2$Ge$_2$, demonstrating essential dissimilarity of electronic states close to the Fermi level which decides the ground state of Eu ions in the compounds. The electronic structure calculations indicate a strong hybridization and a non-integral valence state of Eu in EuNi$_2$P$_2$ whereas Eu-$4f$ states become localized with divalent character in EuNi$_2$Ge$_2$.

\section*{ACKNOWLEDGEMENTS}

We would like to thank Dr. S. Seiro for help with some of the magnetic measurements. This work has been partially supported by the Council of Scientific and Industrial Research, New Delhi (Grant No. 80(0080)/12/ EMR-II).


\begin{thebibliography}{42}

\bibitem{Brandtc5}
N. B. Brandt and V. V. Moshchalkov Adv. Phys. {\bf 33}, 373 (1984).

\bibitem{Monthouxc5}
P. Monthoux, D. Pines and G. G. Lonzarich, Nature {\bf450}, 1177 (2007).

\bibitem{Doniachc5}
 S. Doniach, Physica B {\bf 91}, 231 (1977).

\bibitem{Lacroixc5}
 C. Lacroix and M. Cyrot,  Phys. Rev. B {\bf 20}, 1969 (1979).

\bibitem{Segrec5}
C. U. Segre, M. Croft, J. A. Hodges, V. Murgai, L. C. Gupta and R. D. Parks Phys. Rev. Lett. {\bf49}, 1947 (1982).

\bibitem{Mitsudac5}
A. Mitsuda, Y. Ikeda, N. Ietaka, S. Fukuda, and Y. Isikawa,  J. Magn. Magn. Mater. {\bf310}, 319 (2007).

\bibitem{Hossainc5}
Z. Hossain, C. Geibel, N. Senthilkumaran, M. Deppe, M. Baenitz, F. Schiller, and S. L. Molodtsov,  Phys. Rev. B {\bf69}, 014422 (2004).

\bibitem{Alekseev}
P A Alekseev, K S Nemkovski, J-M Mignot, V N Lazukov, A A Nikonov, A P Menushenkov, A A Yaroslavtsev, R I Bewley, J R Stewart and A V Gribanov, J. Phys.: Condens. Matter {\bf24},  375601 (2012)

\bibitem{Hiranakac5}
Y. Hiranaka, A. Nakamura, M. Hedo, T. Takeuchi, A. Mori, Y. Hirose, K. Mitamura, K. Sugiyama, M. Hagiwara, T. Nakama, and Y.
\={O}nuki, Journal of the Physical Society of Japan {\bf 82}, 083708 (2013).

\bibitem{Guritanu}
 V. Guritanu, S. Seiro, J. Sichelschmidt, N. Caroca-Canales, T. Iizuka, S. Kimura, C. Geibel, and F. Steglich
Phys. Rev. Lett. {\bf 109}, 247207 (2012).

\bibitem{Nagarajanc5}
R. Nagarajan, G. K. Shenoy, L. C. Gupta and E. V. Sampathkumaran, Phys. Rev. B {\bf32}, 2846 (1985).

\bibitem{Geibelc5}
S. Danzenb\"{a}cher, D. V. Vyalikh, Yu. Kucherenko, A. Kade, C. Laubschat, N. Caroca-Canales, C. Krellner, C. Geibel, A. V. Fedorov, D. S. Dessau, R. Follath, W. Eberhardt, and S. L. Molodtsov, Phys. Rev. Lett. {\bf102}, 026403 (2009).

\bibitem{Zahirulc5}
Zahirul Islam, C. Detlefs, C. Song, A. I. Goldman, V. Antropov, B. N. Harmon, S. L. Bud'ko, T. Wiener, P. C. Canfield, D. Wermeille, and K. D. Finkelstein, Phys. Rev. Lett. {\bf 83}, 2817 (1999).

\bibitem{Layekc5}
 S. Layek, V. K. Anand, and Z. Hossain, J. Magn. Magn. Mater {\bf 321}, 3447 (2009).

\bibitem{Rojasc5}
 D. P. Rojas, L. C. J. Pereira, P. Salamakha, E. B. Lopes, J. C. Waerenborgh, L. M. da Silva, and F. G. Gandra, J. Alloys Compd. {\bf 391}, L5 (2005).

\bibitem{Kowalczykc5}
 A. Kowalczyk, M. Pugaczowa-Michalska, and T. Toli$\acute{n}$ski, Phys. Status Solidi B {\bf 242}, 433 (2005).

\bibitem{Mazumdarc5}
 C. Mazumdar, R. Nagarajan, S. K. Dhar, L. C. Gupta, R. Vijayaraghavan, and B. D. Padalia,  Phys. Rev. B {\bf 46}, 9009 (1992).

\bibitem{Pandeyc5}
Abhishek Pandey, Chandan Mazumdar and R Ranganathan, J. Phys.: Condens. Matter {\bf21}, 216002 ((2009).

\bibitem{Tchoulac5}
M.B. Tchoula Tchokont\'{e}, P. de V. du Plessis, D. Kaczorowski, Physica B {\bf 404}, 2992 (2009)

\bibitem{Allenoc5}
 E. Alleno, Z. Hossain, C. Godart, R. Nagarajan and L. C. Gupta, Phys. Rev. B {\bf 52}, 7428 (1995).

\bibitem{Hirstc5}
 L. L. Hirst,  Phys. Kondens. Mater. {\bf 11}, 255 (1970).

\bibitem{Salesc5}
 B. C. Sales, D. K. Wohlleben,  Phys. Rev. Lett. {\bf 35}, 1240 (1975).

\bibitem{Franzc5}
 W. Franz, F. Steglich, W. Zell, D. Wohlleben, and F. Pobell, Phys. Rev. Lett. {\bf 45}, 64 (1980).

\bibitem{Dhar1c5}
S. K. Dhar, R. Nagarajan, S. K. Malik, R. Vijayaraghavan, M. M. Abd-Elmeguid and H. Micklitz, Phys. Rev. B {\bf 29} 5953 {1984}

\bibitem{Dhar2c5}
S. K. Dhar, S. K. Malik, D. Rambabu, and R. Vijayaraghavan, J. Appl. Phys. {\bf53}, 8077 (1982).

\bibitem{Strydomc5}
A. M. Strydom and R. Troc, Solid State Commun. {\bf126}, 207 (2003).

\bibitem{Kittelc5}
C. Kittel, \emph{Introduction to Solid State Physics}, 8th edition (Wiley, New York, 2005).

\bibitem{Goetschc5}
R. J. Goetsch, V. K. Anand, Abhishek Pandey, and D. C. Johnston,  Phys. Rev. B. {\bf 85}, 054517 (2012).


\bibitem{Knoppc5}
G. Knopp,  A. Loidl, R. Caspary,  U. Gottwick,  C. D. Bredl,  H. Spille, F. Steglich and A. P. Murani, Journal  of  Magnetism  and  Magnetic  Materials
 {\bf74}, 341 (1988).

\bibitem{Schoenesc5}
J. Schoenes, C. Sch\"{}nenberger, J. J. M. Franse and A. A. Menovsky, Phys. Rev. B. {\bf 35}, 5375 (1987).

\bibitem{Dasc5}
I. Das and E. V. Sampathkumaran, Phys. Rev. B. {\bf44}, 9711 (1991).

\bibitem{Gignouxc5}
D. Gignoux and J. C. Gomez-Sal, Phys. Rev. B. {\bf30}, 3967 (1984).

\bibitem{Blahac5}
 P. Blaha, K. Schwarz, and G. Madsen,  An Augmented Plane Wave Plus Local Orbitals Program for Calculating Crystal Properties. Vienna University of Technology, Vienna  (2001).

\bibitem{Perdewc5}
 J. P. Perdew, K. Burke, and M. Ernzerhof,  Phys. Rev. Lett. {\bf 77}, 3865 (1996).

\bibitem{Pickettc5}
M. D. Johannes, W. E. Pickett, Phys. Rev. B {\bf72}, 195116 (2005).

\bibitem{Deniszczykc5}
J. Deniszczyk, W. Burian, P.Ma\'{s}lankiewicz, J. Szade, Journal of Alloys and Compounds {\bf442}, 239 (2007).

\bibitem{Felserc5}
C. Felser, S.Cramm, D. Johrendt, A.Mewis, O. Jepsen, G. Hohlneicher, W.Eberhardt and O. K. Andersen, Europhys. Lett., {\bf40}(1), 85 (1997).

\end{thebibliography}
\end{document}